\documentclass[12pt,draftcls,onecolumn]{IEEEtran}
\usepackage{setspace}
\usepackage{clrscode}

\usepackage{amsmath}
\usepackage{amssymb}
\usepackage[dvips]{graphicx}
\usepackage{setspace}
\usepackage{epsfig}

\newcommand{\x}{\mathbf{x}}
\newcommand{\y}{\mathbf{y}}

\newcommand{\n}{\mathbf{n}}

\newcommand{\I}{\mathbf{I}}

\newcommand{\E}{\mathbb{E}}

\newcommand{\Q}{Q^{-1}(\e)}
\newcommand{\dQ}{\dot{Q}^{-1}(\e)}
\newcommand{\ddQ}{\ddot{Q}^{-1}(\e)}

\newcommand{\e}{\epsilon}

\newcommand{\br}{\bar{r}}

\newcommand{\figsize}{0.6}

\newcommand{\tsnr}{{\text{\footnotesize{SNR}}}}

\newcommand{\ssnr}{\text{\scriptsize{SNR}}}

\newtheorem{prop}{Proposition}

\newtheorem{cor}{Corollary}
\newtheorem{lem}{Lemma}

\begin{document}

\title{Throughput Analysis of Buffer-Constrained Wireless Systems in the Finite Blocklength Regime}

\author{\vspace{1cm}
\authorblockN{Mustafa Cenk Gursoy}
\thanks{The author is with the Department of Electrical
Engineering, University of Nebraska-Lincoln, Lincoln, NE, 68588
(e-mail: gursoy@engr.unl.edu).}
\thanks{This work was supported by the National Science Foundation under Grants CCF -- 0546384 (CAREER), CNS -- 0834753, and CCF-0917265.
}}



%


\maketitle

\begin{abstract}
In this paper, wireless systems operating under queueing constraints in the form of limitations on the buffer violation probabilities are considered. The throughput under such constraints is captured by the effective capacity formulation. It is assumed that finite blocklength codes are employed for transmission. Under this assumption, a recent result on the channel coding rate in the finite blocklength regime is incorporated into the analysis and the throughput achieved with such codes in the presence of queueing constraints and decoding errors is identified. Performance of different transmission strategies (e.g., variable-rate, variable-power, and fixed-rate transmissions) is studied. Interactions between the throughput, queueing constraints, coding blocklength, decoding error probabilities, and signal-to-noise ratio are investigated and several conclusions with important practical implications are drawn.
\\


\end{abstract}

\newpage
\setcounter{page}{1}

\begin{spacing}{1.7}
\section{Introduction} \label{sec:intro}

Providing quality of service (QoS) guarantees in the form of limitations on the queueing delays or buffer violation probabilities is essential in many delay-sensitive wireless systems, e.g., voice over
IP (VoIP), and wireless interactive and streaming video applications. Due to the importance of such QoS considerations, it is of significant interest to conduct an analysis and provide predictions for the performance levels of practical systems.
In \cite{dapeng}, effective capacity is proposed as a metric that can be employed to measure the performance
in the presence of statistical QoS limitations. Effective capacity formulation uses the large deviations
theory and incorporates the statistical QoS constraints by capturing the rate of decay of the buffer
occupancy probability for large queue lengths. Hence, effective capacity can be regarded as the maximum
throughput of a system operating under limitations on the buffer violation probability.

Recently, there has been much interest in the analysis of the effective capacity of fading channels (see e.g., \cite{tang-powerrate} -- \cite{imperfect}) in order to identify the performance of wireless systems operating under statistical queueing constraints. However, in almost all prior studies, the service rates of the queueing model (or equivalently the instantaneous transmission rates over the wireless channel) are assumed to be equal to the instantaneous capacity values although channel coding is performed using a finite block of symbols. Moreover, transmissions are assumed to be reliable with no decoding errors. However, it is important to note that error-free communication at the rate of channel capacity is generally attained as the codeword length increases without bound. Therefore, when finite blocklength codes are employed, transmission is necessarily performed in the presence of decoding errors and possibly at rates less than the channel capacity in order to have high reliability or equivalently low error probability.

In \cite{negi-goel} and \cite{goel-negi}, Negi and  Goel addressed these considerations. They studied queueing and coding jointly and took explicitly into account decoding errors by considering the random coding exponents of error probabilities for rates less than the instantaneous channel capacity. For instance, in \cite{negi-goel}, they analyzed the maximization of the joint exponent of the decoding error and delay violation probability through the appropriate choice of the transmission rate for given delay bound and constant arrival rate.

In this paper, we also depart from the idealistic assumptions of communicating arbitrarily reliably at channel capacity but follow an approach different from that of \cite{negi-goel} and \cite{goel-negi}.  We consider channel coding rates achievable with finite blocklength codes, and incorporate the decoding error probabilities and possible retransmission scenarios into the effective capacity formulation. This analysis is facilitated mainly by the recent results of Polyanskiy, Poor, and Verd\'u in \cite{yury} where the authors identified an approximate maximal achievable rate expression for a given error probability in the finite blocklength regime. This expression can be regarded as a second-order asymptotic approximation of the channel coding rate at large but finite blocklength values. We note that \cite{laneman} and \cite{valenti} also studied channel coding and achievable error probabilities at finite blocklengths by analyzing the mutual information density and its statistics. In \cite{valenti}, an outage analysis is performed by using the distribution of the mutual information density. In \cite{pengwu}, a similar outage formulation is used to determine the optimal physical-layer reliability and to identify the maximum ARQ throughput. On the other hand, neither of the above-mentioned papers have investigated the throughput in the finite blocklength regime when the systems operate under buffer constraints.

Our contributions in this paper can be summarized as follows. We first determine the effective throughput in the finite blocklength regime under constraints on the buffer violation probability. Subsequently, we study the performance of different transmission strategies. Initially, we consider a scenario in which the transmission rate is varied with the fading realizations while the error probability is kept fixed. The optimal error probability that maximizes the throughput is shown to be unique. We analyze the impact of the power adaptations. Then, we investigate the case in which transmission rate is fixed and error probability varies over different transmission blocks. Through numerical results, we analyze the interactions between the throughput, queueing constraints, error probabilities, blocklength, signal-to-noise ratio, and different transmission strategies.

The remainder of the paper is organized as follows. Section \ref{sec:channelmodel} describes the fading channel model. In Section \ref{sec:effcap-background}, we provide preliminaries on the effective capacity as a measure of the throughput under statistical QoS constraints. In Section \ref{sec:effcap}, we provide our results on the effective throughput in the finite blocklength regime. We conclude in Section \ref{sec:conclusion}. Several proofs are relegated to the Appendix.

\section{Channel Model} \label{sec:channelmodel}

We consider a frequency-flat channel model, and assume that the fading coefficients stay fixed for a block of $m$ symbols and then change independently for the following block. Under this block-fading assumption, the channel input-output relation in one coherence block can be expressed as
\vspace{-.1cm}
\begin{gather}
\y = h \x + \n
\end{gather}
where $\x$ and $\y$ are the $m$-dimensional, complex, channel input and output vectors, respectively. The input is subject to an average power constraint, i.e., $\E\{\|\x\|^2\}\le mP$. $h$ is the complex-valued fading coefficient with finite second moment, i.e., $\E\{|h|^2\} < \infty$. We assume that both the receiver and transmitter have perfect channel side information (CSI) and hence perfectly know the instantaneous realizations of the fading coefficients. However, the assumption of perfect CSI at the transmitter is relaxed in Section \ref{subsec:fixed-rate}. Finally, $\n$ represents the Gaussian noise vector whose components are independent and identically distributed (i.i.d.), complex, circularly symmetric, Gaussian random variables with mean zero and variance $N_0$, i.e., $\n \sim \mathcal{CN} (0, N_0 \I_m)$ where $\I_m$ denotes the $m \times m$ identity matrix.

\section{Throughput under Statistical Queueing Constraints} \label{sec:effcap-background}

In \cite{dapeng}, Wu and
Negi  defined the effective capacity as the maximum
constant arrival rate that a given service process can support in
order to guarantee a statistical QoS requirement specified by the
QoS exponent $\theta$ \footnote{For time-varying arrival rates,
effective capacity specifies the effective bandwidth of the arrival
process that can be supported by the channel.}. If we define $Q$ as
the stationary queue length, then $\theta$ is the decay rate of the
tail of the distribution of the queue length $Q$:
\begin{equation}
\lim_{q \to \infty} \frac{\log P(Q \ge q)}{q} = -\theta.
\end{equation}
Therefore, for large $q_{\max}$, we have the following approximation
for the buffer violation probability: $$P(Q \ge q_{\max}) \approx
e^{-\theta q_{\max}}.$$ Hence, while larger $\theta$ corresponds to
more strict QoS constraints, smaller $\theta$ implies looser QoS
guarantees. Similarly, if $D$ denotes the steady-state delay
experienced in the buffer, then $P(D \ge d_{\max}) \approx
e^{-\theta \delta d_{\max}}$ for large $d_{\max}$, where $\delta$ is
determined by the arrival and service processes
\cite{Tang-crosslayer2}. Therefore, effective capacity formulation provides the maximum constant arrival rates that can be supported by the time-varying wireless channel under the queue length constraint $P(Q \ge q_{\max}) \le
e^{-\theta q_{max}}$ for large $q_{max}$ or the delay constraint $P(D \ge d_{\max}) \le
e^{-\theta \delta d_{\max}}$ for large $d_{\max}$. Since the average
arrival rate is equal to the average departure rate when the queue
is in steady-state \cite{ChangZajic}, effective capacity can also be
seen as the maximum throughput in the presence of such constraints.

The effective capacity is given by (\cite{dapeng}, \cite{chang}, \cite{cschang})
\begin{gather}
R_E = -\lim_{t\rightarrow\infty}\frac{1}{\theta
t}\log_e{\mathbb{E}\{e^{-\theta S[t]}\}}
\end{gather}
where $S[t] = \sum_{i=1}^{t}R_i$ is the time-accumulated service
process and $\{R_i, i=1,2,\ldots\}$ denotes the discrete-time
stationary and ergodic stochastic service process. We would like to note that in the remainder of the paper, we will refer to $R_E$ as the effective rate rather than the effective capacity since $R_E$ in our setup is the throughput when the service rates are equal to the approximate channel coding rates in the finite blocklength regime.

\vspace{-.5cm}
\section{Effective Throughput with Finite Blocklength Codes} \label{sec:effcap}

In \cite{yury}, the authors have studied the channel coding rate in the finite blocklength regime. For general classes of channels, they have obtained new achievability and converse bounds on the coding rate for a given finite blocklength and error probability. In particular, for the real, additive white Gaussian noise (AWGN) channel, the transmission rate (in bits per $m$ channel uses) with error probability $0 < \e < 1$, signal-to-noise ratio (\tsnr), and coding blocklength $m$ is shown to have the following asymptotic expression \cite[Theorem 54]{yury}:
\vspace{-.5cm}
\begin{align}\label{eq:codingrateactual}
r = \frac{m}{2}\log_2(1+\ssnr) - \sqrt{\frac{m}{2} \left( 1- \frac{1}{(\ssnr + 1)^2} \right)} &Q^{-1}(\e) \log_2 e + O(\log m)
\end{align}
where $Q(x) = \int_{x}^\infty \frac{1}{\sqrt{2\pi}} \, e^{-t^2/2}\, dt$ is the Gaussian $Q$-function. Denoting the rate in bits per channel use by $\br$, we can write
\begin{align}
\br =  \frac{r}{m} &= \frac{1}{2}\log_2(1+\ssnr) - \sqrt{\frac{1}{2m} \left( 1- \frac{1}{(\ssnr + 1)^2} \right)} Q^{-1}(\e) \log_2 e + \frac{O(\log m)}{m}
\\
&\approx \frac{1}{2}\log_2(1+\ssnr) - \sqrt{\frac{1}{2m} \left( 1- \frac{1}{(\ssnr + 1)^2} \right)} Q^{-1}(\e) \log_2 e \label{eq:codingrate}
\end{align}
where the approximation is accurate for sufficiently large $m$.
Note that the above results are for the AWGN channel with real input and real output.

In this paper, we consider a fading Gaussian channel model with complex-valued input and output, and assume that channel coding is performed in each coherence interval of $m$ symbols, during which the fading stays fixed.  Under these assumptions, coding over a fading Gaussian channel can be seen as coding over a real Gaussian channel (with a certain channel gain) using a coding blocklength of $2m$.  The following arguments provide a detailed description of this approach. Knowing the  channel fading coefficient $h$, the receiver can multiply the received signal with $e^{-j \theta_h}$, where $\theta_h$ is the phase of $h$, and obtain\footnote{Note that multiplication of the channel output with the $e^{-j \theta_h}$ just rotates the output, is a reversible operation, and hence does not lead to any loss of information.}
\begin{align}
\tilde{\y} =   \tilde{\y}_r  + j \tilde{\y}_i = \y e^{-j\theta_h} &= |h| \x + \tilde{\n}
= |h| \x_r + \tilde{\n}_r + j (|h| \x_i + \tilde{\n}_i)
\end{align}
where $\tilde{\y}_r$, $\x_r$, $\tilde{\n}_r$ and $\tilde{\y}_i$, $\x_i$, $\tilde{\n}_i$ denote the real and imaginary components, respectively, of the output vector $\tilde{\y}$, input vector $\x$, and noise vector $\tilde{\n}$.  It can be easily verified that $\tilde{\n} = \n e^{-j \theta_h}$ has the same statistics as $\n$ and hence $\tilde{\n} \sim \mathcal{CN} (0, N_0 \I_m)$. Now, the above channel input-output relation can also be written as
\begin{gather}
[\tilde{\y}_r \,\, \tilde{\y}_i]= |h| \, [\x_r \,\, \x_i] + [\tilde{\n}_r \,\, \tilde{\n}_i]
\end{gather}
where $[\tilde{\y}_r \,\, \tilde{\y}_i]$ denotes the vector formed by concatenating $\tilde{\y}_r$ and $\tilde{\y}_i$. Since the real and imaginary components are $m$-dimensional vectors, the above channel model is a real Gaussian channel with $2m$ dimensional input and output and with channel gain $|h|$.
Note that the real and imaginary noise components $\tilde{\n}_r$ and  $\tilde{\n}_i$ are independent due to the assumption of the circular symmetry of the additive complex Gaussian noise. For this channel, the coding rate (in bits per $m$ channel uses) in the $i^{\text{th}}$ block achieved with  block error probability $\e$ is
\vspace{-.5cm}
\begin{align}
r_i = &m\log_2(1+\ssnr|h_i|^2) - \sqrt{m \left( 1- \frac{1}{(\ssnr|h_i|^2 + 1)^2} \right)} Q^{-1}(\e) \log_2 e  + O(\log 2m) \label{eq:codingratef-m-uses}
\end{align}
where $h_i$ denotes the fading coefficient in the $i^{\text{th}}$ block. Note that  the expression in (\ref{eq:codingratef-m-uses}) is obtained from that in (\ref{eq:codingrateactual}) by replacing $m$ with $2m$, and $\tsnr$ with $\tsnr |h_i|^2 = \frac{P}{N_0} |h_i|^2$, which is the received signal-to-noise ratio in the $i^{\text{th}}$ block. Now, the normalized rate in bits per channel use is approximately
\begin{gather}
\br_i = \frac{r_i}{m} =  \log_2(1+\ssnr|h_i|^2) - \sqrt{\frac{1}{m} \left( 1- \frac{1}{(\ssnr|h_i|^2 + 1)^2} \right)} Q^{-1}(\e) \log_2 e  \label{eq:codingratef}
\end{gather}
for large enough $m$ for which $\frac{O(\log 2m)}{m}$ is negligible. Henceforth, we assume that the instantaneous transmission rate in each coherence block of the fading channel is given by the expression in (\ref{eq:codingratef}). Since the block error rate is $\e$,  this rate is attained with probability $1-\e$. We assume that the receiver reliably detects the errors, employs a simple ARQ mechanism and sends a negative acknowledgement requesting the retransmission of the message in case of an erroneous reception. Therefore, the data rate is effectively zero when error occurs. Under this assumption,  the service rate (in bits per $m$ channel uses) in each block is
\begin{align}
R_i = \left\{
\begin{array}{ll}
0 & \text{with prob. } \e
\\
m\br_i & \text{with prob. } (1-\e)
\end{array}\right.. \label{eq:servicerate}
\end{align}
With the above service rate characterization, we immediately obtain the following expression for the effective rate.

\begin{prop} \label{prop:effrate}
The effective rate (in bits per channel use) at a given $\tsnr$, error probability $\e$, blocklength $m$, and QoS exponent $\theta$ is
\vspace{-.5cm}
\begin{gather} \label{eq:effthr}
R_E(\theta) = -\frac{1}{m \theta} \log_e \E_{|h|^2}\left\{\e + (1-\e) e^{-\theta m \br} \right\}
\end{gather}
where $\br$ is given in (\ref{eq:codingratef}) and the expectation is with respect to $|h|^2$.
\end{prop}

\emph{Proof:} We first note that the service rate $\{R_i\}$ is an i.i.d. process due to the facts that the fading process is i.i.d. in different blocks and the noise is an i.i.d. process leading to the independence of error events in different blocks. Now, we have
\begin{align}
R_E(\theta) &= -\lim_{t\rightarrow\infty}\frac{1}{\theta
t}\log_e{\mathbb{E}\{e^{-\theta S[t]}\}}
\\
&= -\lim_{t\rightarrow\infty}\frac{1}{\theta
t}\log_e{\mathbb{E}\{e^{-\theta \sum_{i=1}^{t}R_i}\}}
\\
&= -\lim_{t\rightarrow\infty}\frac{1}{\theta
t}\log_e{\mathbb{E}\left\{\prod_{i=1}^{t} e^{-\theta R_i}\right\}}
\\
&= -\lim_{t\rightarrow\infty}\frac{1}{\theta
t}\log_e{\prod_{i=1}^{t}\mathbb{E}\left\{ e^{-\theta R_i}\right\}} \label{eq:effthrproof1}
\\
&= -\lim_{t\rightarrow\infty}\frac{1}{\theta
t}\log_e{\left(\mathbb{E}\left\{ e^{-\theta R_i}\right\}\right)^t} \label{eq:effthrproof2}
\\
&= -\lim_{t\rightarrow\infty}\frac{1}{\theta
t} t \log_e{\mathbb{E}\left\{ e^{-\theta R_i}\right\}} \label{eq:effthrproof3}
\\
&= -\frac{1}{\theta} \log_e{\mathbb{E}\left\{ e^{-\theta R_i}\right\}} \label{eq:effthrproof4}
\\
&= -\frac{1}{\theta} \log_e{\mathbb{E}_{|h|^2}\left\{ \e + (1-\e)e^{-\theta m\br_i} \right\}} \label{eq:effthrproof5}
\end{align}
Above, (\ref{eq:effthrproof1}) follows from the independence of the service process and (\ref{eq:effthrproof2}) is due to its being identically distributed. The expression inside the expectation in (\ref{eq:effthrproof5}) is obtained by evaluating the expected value of $e^{-\theta R_i}$ for fixed $|h|^2$. Finally, (\ref{eq:effthr}) is obtained by normalizing (\ref{eq:effthrproof5}) by $m$ to have the effective rate in the units of bits per channel use, and by dropping the time index $i$. \hfill $\square$

Note that the effective rate is a function of the QoS exponent $\theta$, blocklength $m$, signal-to-noise ratio $\tsnr$ and error probability $\e$. Since we assume that coding is performed in each coherence interval, the blocklength $m$ is determined by the statistics of the fading process. The value of $\theta$ can be dictated by the application requirements and $\tsnr$ depends on the power budget. Given the values of these parameters, the remaining parameter $\e$ can be optimized to maximize the throughput. Note that large $\e$ implies that the transmitter attempts to transmit the data at a high rate but at the risk of more frequent errors and hence retransmissions. On the other hand, if $\e$ is small, the instantaneous transmission rate is low but the reliability of the transmissions is high. The following result shows that the optimal $\e$ is unique.

\begin{prop} \label{prop:convexity}
Assume that the values of $m$, $\theta > 0$, and $\tsnr > 0$ are fixed. Then, the function
\begin{gather} \label{eq:functioninsidelog}
\Psi(\e) = \E_{|h|^2}\left\{\e + (1-\e)e^{-\theta m\br}\right\}
\end{gather}
is strictly convex in $\e$ and therefore the optimal value of $\e$ that minimizes this function or equivalently maximizes the effective rate in (\ref{eq:effthr}) is unique.
\end{prop}

\emph{Proof}: See Appendix \ref{app:convexity}.

Note that the convexity result indicates that the optimal error probability $\e^*$ can be easily found using standard convex optimization methods. The analysis and the resulting $\e^*$ provide guidelines on the design of the channel codes and their strength. Note further that the above result is shown for the case in which $\theta > 0$. If there are no QoS constraints and hence $\theta = 0$, then we have the following corollary to Proposition \ref{prop:effrate}.
\begin{cor}
When $\theta = 0$, the effective capacity becomes
\begin{gather} \label{eq:R_E(0)}
R_E(0) = \lim_{\theta \to 0} R_E(\theta) = (1-\e) \E_{|h|^2}\{\br\}
\end{gather}
where $\br$ is given in (\ref{eq:codingratef}).
\end{cor}

Note that the $R_E(0)$ is the average transmission rate averaged over the fading states. Below, we show that $R_E(0)$ is a strictly concave function of $\e$.

\begin{prop}\label{prop:convexity2}
Assume that the values of $m$, and $\tsnr > 0$ are fixed. Then, the function
\begin{gather}
R_E(0) = (1-\e) \E_{|h|^2}\{\br\}
\end{gather}
is strictly concave in $\e$ and therefore the optimal value of $\e$ that maximizes this effective rate is unique.
\end{prop}

\emph{Proof}: See Appendix \ref{app:convexity2}.

Next, we provide numerical examples to illustrate the results. Although the preceding analysis is applicable to any fading distribution with finite power, we consider a Rayleigh fading channel in the numerical analysis, and assume that the fading power $z = |h|^2$ is exponentially distributed with unit mean (i.e., has the probability density function $f_z(z) = e^{-z}$).

\begin{figure}
\begin{center}
\includegraphics[width=\figsize\textwidth]{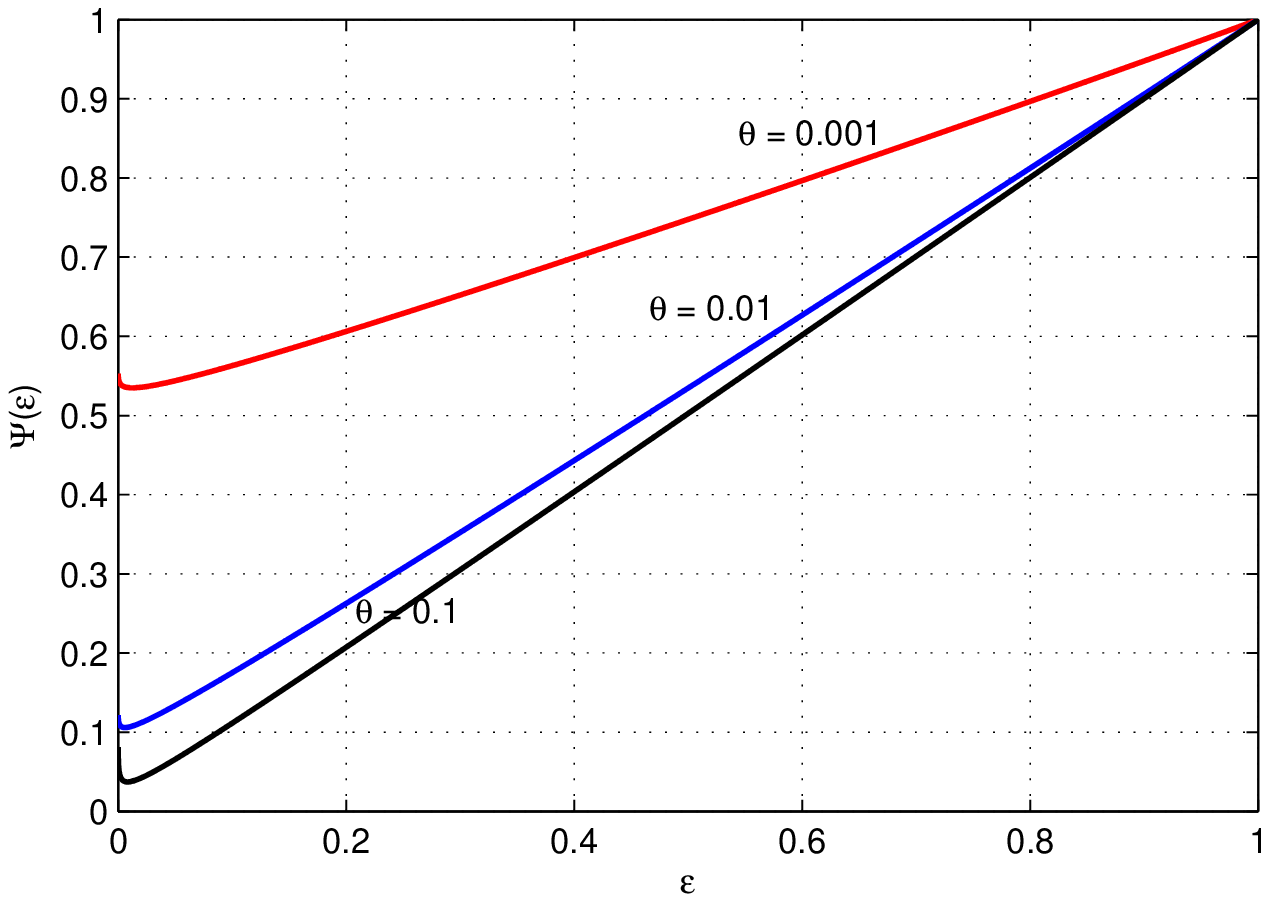}
\caption{The function $\Psi(\e)$ vs. the error probability $\e$ in the Rayleigh fading channel. $\tsnr = 0$ dB and the blocklength is $m = 1000$.}\label{fig:psi-eps}
\end{center}
\end{figure}

In Figure \ref{fig:psi-eps}, we plot $\Psi(\e)= \E_{|h|^2}\left\{\e + (1-\e)e^{-\theta m\br}\right\}$ as a function of the error probability $\e$ in the Rayleigh fading channel. In the figure, $\tsnr = 0$ dB and the blocklength $m = 1000$. We provide curves for different values of the QoS exponent $\theta > 0$. In all cases, we immediately observe the strict convexity of the curves, confirming the result in Proposition \ref{prop:convexity}. Indeed, the optimal error probabilities that minimize $\Psi(\e)$ are unique and are equal to $\e^* = 0.0127, 0.0061, 0.0084$ for $\theta = 0.001, 0.01, 0.1$, respectively.

\begin{figure}
\begin{center}
\includegraphics[width=\figsize\textwidth]{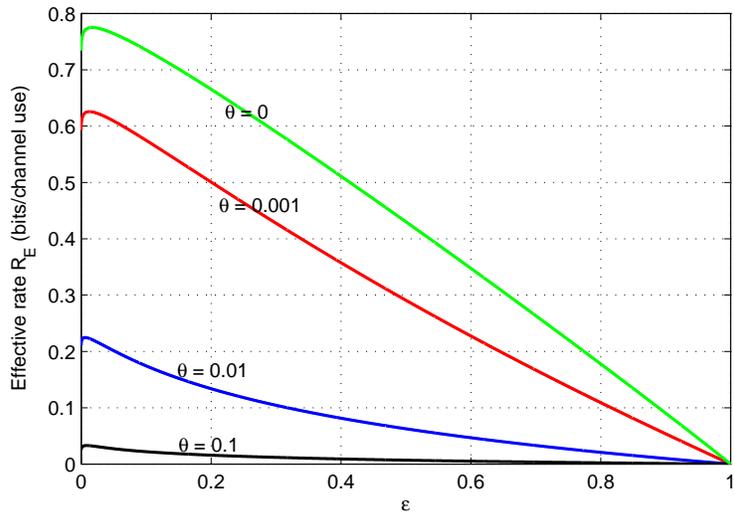}
\caption{Effective rate $R_E$ vs. the error probability $\e$ in the Rayleigh fading channel. $\tsnr = 0$ dB and the blocklength is $m = 1000$.}\label{fig:R_E-eps}
\end{center}
\end{figure}

\begin{figure}
\begin{center}
\includegraphics[width=\figsize\textwidth]{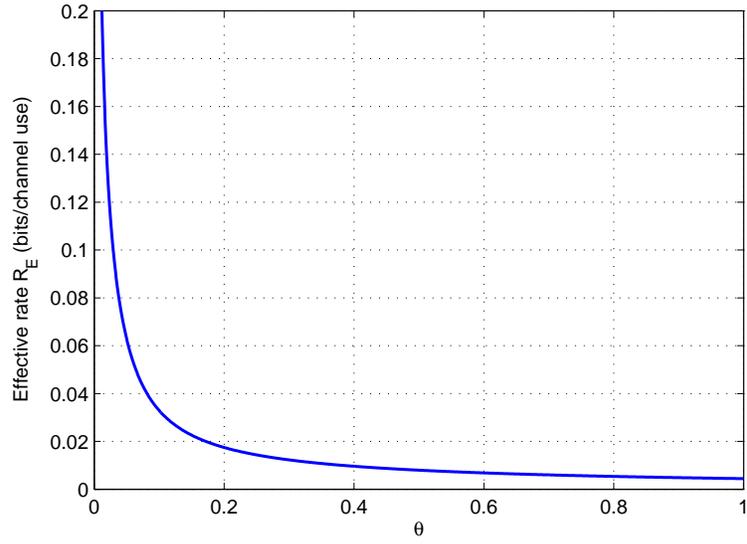}
\caption{The optimal effective rate $R_E$ vs. QoS exponent $\theta$ in the Rayleigh fading channel. $\tsnr = 0$ dB and the blocklength is $m = 1000$.}\label{fig:R_E-theta}
\end{center}
\end{figure}

In Fig. \ref{fig:R_E-eps}, we plot the effective rate in (\ref{eq:effthr}) as a function of the error probability $\e$. The other parameters are the same as in Fig. \ref{fig:psi-eps}. Notice that we have also included in this figure the throughput curve for the case in which $\theta = 0$. Note that if $\theta = 0$, the system does not have any queueing constraints. In Proposition \ref{prop:convexity2},  we have shown that $R_E(0)$ is a strictly concave function of $\e$ and the optimal $\e^*$ that maximizes $R_E(0)$  is unique. The strict concavity is observed in Fig. \ref{fig:R_E-eps}. The optimal value of the error probability in the case of $\theta = 0$ is $\e^* = 0.0171$. For $\theta > 0$, the effective rate curves are not necessarily concave. In Fig. \ref{fig:R_E-eps},  we observe that these curves are quasiconcave and, as predicted by Proposition \ref{prop:effrate}, they are maximized at a unique $\e^*$. The optimal error probabilities for the cases in which $\theta > 0$ are equal to the same ones obtained in Fig. \ref{fig:psi-eps}. At the optimal error probabilities, the maximum effective rate values are $R_E = 0.7750,0.6256,0.2246,0.0329$ bits/channel use for $\theta = 0, 0.001, 0.01, 0.1$, respectively. Note that increasing $\theta$ leads to more stringent QoS constraints, and we observe that the effective rate and hence the effective throughput diminishes as $\theta$ increases. This trend is also clearly seen in Fig. \ref{fig:R_E-theta} where we plot the maximum effective rate values (i.e., effective rate at the optimal error probability $\e^*$) as a function of $\theta$.

\begin{figure}
\begin{center}
\includegraphics[width=\figsize\textwidth]{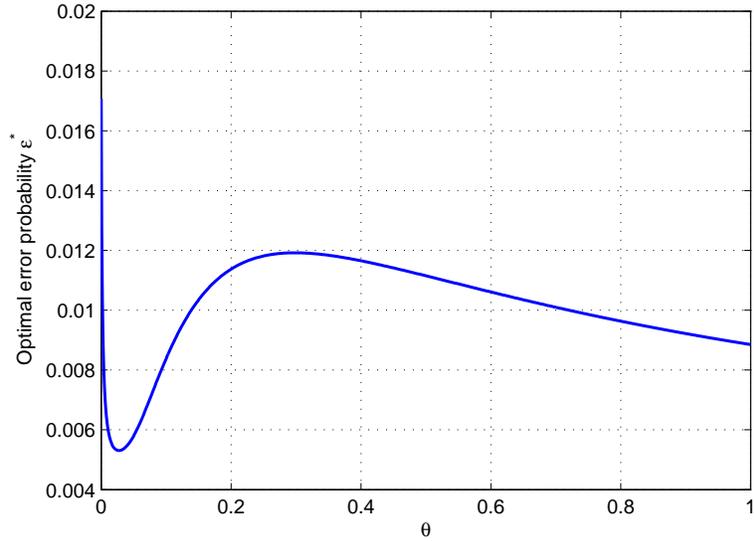}
\caption{The optimal error probability $\e^*$ vs. QoS exponent $\theta$ in the Rayleigh fading channel. $\tsnr = 0$ dB and the blocklength is $m = 1000$.}\label{fig:eps-theta}
\end{center}
\end{figure}

Another interesting analysis is the behavior of $\e^*$ as a function of $\theta$. This is depicted in Fig. \ref{fig:eps-theta}. Here, we observe that as $\theta$ increases and therefore the QoS limitations become more stringent, the value of $\e^*$ initially decreases sharply. Hence, the transmitter opts for more reliable but low-rate transmissions. On the other hand, as $\theta$ increases beyond approximately 0.028, the trend reverses and $\e^*$ starts to increase. The transmitter increases the transmission rate at the cost of increased $\e^*$ and hence more retransmissions. When $\theta$ exceeds 0.298, $\e^*$ starts decreasing again.
Note that for high values of $\theta$, the effective rate is small. This small effective rate can be supported by low-rate transmissions. Hence, when $\theta$ is high beyond a threshold, the transmitter chooses to transmit at low rates and keep the error probability and the number of retransmissions low as well.

\begin{figure}
\begin{center}
\includegraphics[width=\figsize\textwidth]{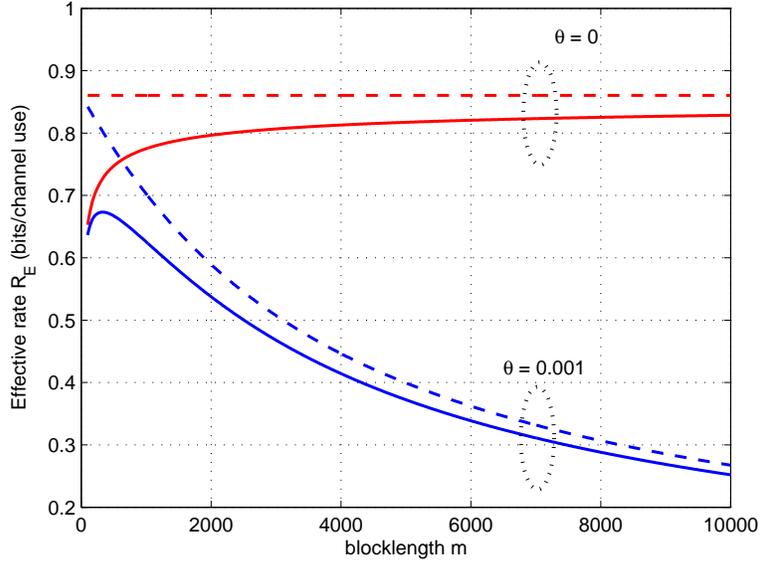}
\caption{The optimal effective rate $R_E$ vs. the blocklength $m$ in the Rayleigh fading channel. $\tsnr = 0$ dB and the QoS exponent is $\theta = 0.001$. Dashed curves correspond to the effective rate of the ideal model in which the service rate is equal to the instantaneous channel capacity and error probability is zero.}\label{fig:R_E-n}
\end{center}
\end{figure}

In Fig. \ref{fig:R_E-n}, we plot the effective rate as a function of the blocklength $m$ for $\theta = 0$ and $\theta = 0.001$. The solid-lined curves correspond to the effective rate in (\ref{eq:effthr}) optimized over $\e$. The dashed curves correspond to the effective rate of the ideal model in which the service rate is equal to the instantaneous capacity, i.e.,
\begin{gather}
\br = \log_2(1 + \ssnr |h|^2),
\end{gather}
and the error probability is assumed to be zero, i.e., $\e = 0$.
Here, we have interesting observations. When $\theta = 0$ and the ideal model is considered, then the effective rate is $R_E(0) = \E_{|h|^2}\{\log_2(1 + \ssnr |h|^2)\}$, which is the ergodic capacity of the fading channel and is clearly independent of the blocklength. On the other hand, if the service rate is given by $\br$ in (\ref{eq:codingratef}), the effective rate $R_E(0) = (1-\e) \E_{|h|^2}\{\br\}$ increases with blocklength $m$ as seen in Fig. \ref{fig:R_E-n}. In the  presence of QoS constraints, i.e., when $\theta > 0$, we have stark differences. Under the idealistic assumption of transmitting at the instantaneous capacity with no errors, we see from the behavior of the dashed curve for $\theta = 0.001$ that effective rate decreases with increasing $m$. The reason is that since $m$ is the coherence duration over which the fading state remains fixed, larger $m$ corresponds to slower fading and slow fading is detrimental for buffer-constrained systems. In a slow-fading scenario, deep-fading can be persistent causing long durations of low rate transmissions leading to buffer overflows. In the finite blocklength regime, as seen in the behavior of the solid-lined curve of the case of $\theta = 0.001$, there is a certain tradeoff. Initially, increasing $m$ improves the performance as this allows the system to perform transmissions with longer codewords and to have higher transmission rates. However, if $m$ increases beyond a threshold, slowness of the fading starts to degrade the performance.

In all cases in Fig. \ref{fig:R_E-n}, the gap between the dashed and solid-lined curves diminishes as $m$ increases since the idealistic model becomes more accurate. On the other hand, for moderate values of $m$ (e.g., when $m < 2000$), the idealistic assumptions lead to significant overestimations of the performance.

\begin{figure}
\begin{center}
\includegraphics[width=\figsize\textwidth]{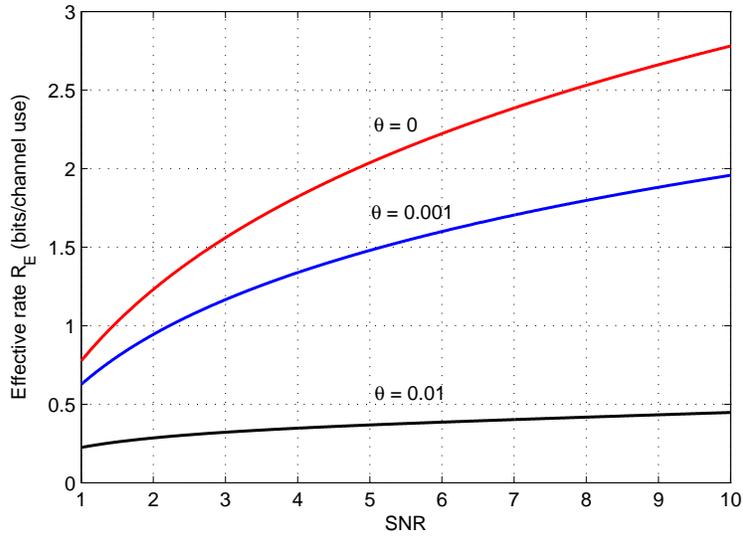}
\caption{The optimal effective rate $R_E$ vs. signal-to-noise ratio (SNR) in the Rayleigh fading channel. The blocklength is $m = 1000$.}\label{fig:R_E-snr}
\end{center}
\end{figure}

\begin{figure}
\begin{center}
\includegraphics[width=\figsize\textwidth]{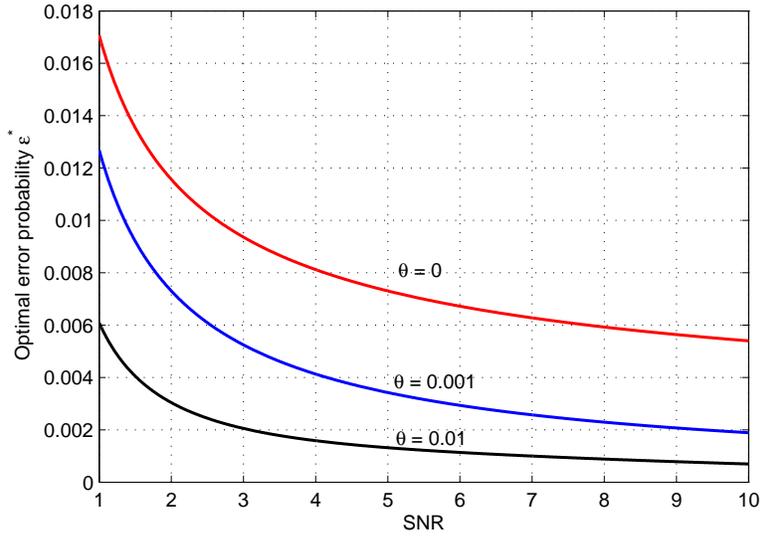}
\caption{The optimal error probability $\e^*$ vs. signal-to-noise ratio (SNR). The blocklength is $m = 1000$.}\label{fig:eps-snr}
\end{center}
\end{figure}

Finally, we provide numerical results for the optimal effective rate and optimal error probability as a function of $\tsnr$ in Figs. \ref{fig:R_E-snr} and \ref{fig:eps-snr}, respectively, for $\theta = 0, 0.001,$ and $0.01$. We see that, for fixed $\theta$, increasing the $\tsnr$ improves the throughput and also the reliability of the transmissions by lowering the error probabilities.

\subsection{The Impact of Power Adaptation} \label{subsec:power}

Heretofore, we have considered the scenario where the transmitter knows the fading coefficients $\{h_i\}$ and performs variable-rate transmission with the same average power $P$ in each coherence block of $m$ channel uses. In this section, we investigate the gains achieved by varying the transmission power as well with respect to fading. Let us denote the power adaptation normalized by the noise  power by $\mu(\tsnr, \theta, |h|^2)$. With this adaptation policy, the transmission rate is
\begin{gather}
\br =  \log_2(1+\mu(\tsnr, \theta, |h|^2)|h|^2) - \sqrt{\frac{1}{m} \left( 1- \frac{1}{(\mu(\tsnr, \theta, |h|^2)|h|^2 + 1)^2} \right)} Q^{-1}(\e) \log_2 e  \label{eq:codingratef-pa}
\end{gather}
which is obtained by replacing $\tsnr$ with $\mu(\tsnr, \theta, |h|^2)$ in  (\ref{eq:codingratef}). Finding the optimal power adaptation policy that maximizes $\br$ or the effective rate $
R_E(\theta) = -\frac{1}{m \theta} \log_e \E_{|h|^2}\left\{\e + (1-\e) e^{-\theta m \br} \right\}$ is in general a difficult task due to the facts that both the first and second terms on the right-hand side of (\ref{eq:codingratef-pa}) are concave functions. Hence, $\br$ is neither concave or convex. For this reason, we resort to suboptimal strategies. One viable policy, $\mu^*$,  is the one that maximizes the effective rate when the service process is assumed to be equal to the instantaneous capacity $\log(1 + \mu(\tsnr, \theta, |h|^2)|h|^2)$ with zero error probability, i.e.,
\begin{gather}
 \mu^*(\ssnr, \theta, |h|^2)\}= \arg \max_{E_{|h|^2}\{\mu(\ssnr, \theta, |h|^2)\} \le \ssnr} -\frac{1}{m \theta} \log_e \E_{|h|^2}\left\{e^{-\theta m \log_2(1 + \mu(\tsnr, \theta, |h|^2)|h|^2)} \right\}.
\end{gather}
$\mu^*$ is derived in \cite{tang-powerrate} and is given by
\begin{align}
\mu^*(\ssnr, \theta, |h|^2)\} =  \left\{
\begin{array}{ll}
\frac{1}{\alpha^{\frac{1}{\beta + 1}} \, (|h|^2)^{\frac{\beta}{\beta + 1}}} - \frac{1}{|h|^2} & |h|^2 \ge \alpha
\\
0 & |h|^2 < \alpha
\end{array}\right.
\end{align}
where $\beta = \frac{\theta m}{\log_e 2}$ and $\alpha$ is chosen such that the average long-term signal-to-noise ratio constraint, \\ $E_{|h|^2}\{\mu(\tsnr, \theta, |h|^2)\} \le \tsnr$ is satisfied with equality. Note that this policy is close to the optimal one when the blocklength is large and hence $\br$ is close to $\log(1 + \mu(\tsnr, \theta, |h|^2)|h|^2)$ and $\e$ is close to zero.

\begin{figure}
\begin{center}
\includegraphics[width=\figsize\textwidth]{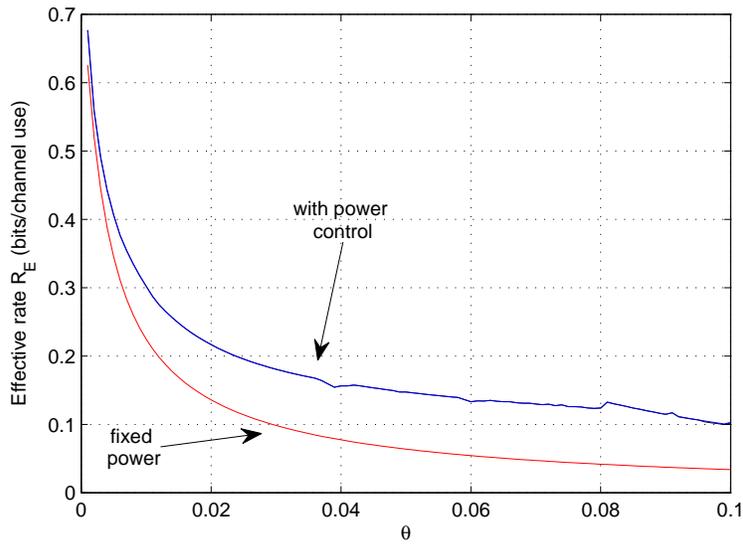}
\caption{The optimal effective rate $R_E$ vs. QoS exponent $\theta$ in the Rayleigh fading channel with and without power control. $\tsnr = 0$ dB and the blocklength is $m = 1000$.}\label{fig:R_E-theta-var-3}
\end{center}
\end{figure}

In Fig. \ref{fig:R_E-theta-var-3}, the optimal effective rate is plotted as a function of $\theta$ for both fixed- and variable-power cases. In the fixed power case, $\tsnr = 0$ dB in each coherence block. When power adaptation is employed, signal-to-noise ratio $\mu(\tsnr, \theta, |h|^2)$ varies in each block while satisfying $\E_{|h|^2}\{\mu(\tsnr, \theta, |h|^2)\} \le \tsnr = 0$ dB. The improved performance with power control is observed in the figure.

\subsection{Fixed-Rate Transmissions} \label{subsec:fixed-rate}

The analysis above has assumed that the transmitter has perfect knowledge of the fading coefficients and can perform variable-rate and/or variable-power transmissions in each coherence block. On the other hand, it is practically interesting to consider cases in which the transmitter does not know the channel and send the information at a fixed rate. Additionally, the transmitter may prefer fixed-rate transmissions, even when it knows the channel, due to complexities in varying the transmission rate for each block. Motivated by these considerations, we assume in this section that the transmitter sends the information at the fixed rate $\br_f$. Under this assumption, error probability $\e$ varies with the fading realizations. The analysis in the previous sections have, on the other hand, considered the scenarios in which the error probability is fixed for all channel states.

From (\ref{eq:codingratef}), which provides the fundamental tradeoff between the rate and error probability in the finite blocklength regime, we can easily see that the error probability for fixed $\br_f$ is
\begin{align}
\e = Q\left( \frac{\log_2(1+\ssnr|h|^2) - \br_f} {\sqrt{\frac{1}{m} \left( 1- \frac{1}{(\ssnr|h|^2 + 1)^2} \right)} \log_2 e}\right).
\end{align}
Note that $\e$ is a function of the fading magnitude $|h|$, signal-to-noise ratio $\tsnr$, and blocklength $m$. The service rate (in bits per $m$ channel uses) is now
\begin{align}
R_i = \left\{
\begin{array}{ll}
0 & \text{with prob. } \e = Q\left( \frac{\log_2(1+\ssnr|h|^2) - \br_f} {\sqrt{\frac{1}{m} \left( 1- \frac{1}{(\ssnr|h|^2 + 1)^2} \right)} \log_2 e}\right)
\\
m\br_f & \text{with prob. } (1-\e)
\end{array}\right.. \label{eq:servicerate-f}
\end{align}
It can also be immediately seen that
for given $\tsnr$, blocklength $m$, QoS exponent $\theta$, and fixed-rate $r_f$, the the effective rate in bits per channel use is
\begin{gather} \label{eq:effth-f}
R_E(\theta) = -\frac{1}{m \theta} \log_e \E_{|h|^2}\left\{\e + (1-\e) e^{-\theta m \br_f} \right\}
\end{gather}
which is essentially the same as in (\ref{eq:effthr}). The only difference is that we now have the rate fixed and error probability varying. Similarly, when $\theta = 0$, we have
\begin{gather} \label{eq:R_E(0)-fixed}
R_E(0) = \E_{|h|^2}\{(1-\e) \br_f\} = (1 - \E_{|h|^2}\{\e\}) \br_f = \left(1 - \E_{|h|^2}\left\{Q\left( \frac{\log_2(1+\ssnr|h|^2) - \br_f} {\sqrt{\frac{1}{m} \left( 1- \frac{1}{(\ssnr|h|^2 + 1)^2} \right)} \log_2 e}\right)\right\}\right) \br_f.
\end{gather}
It is instructive to investigate what is obtained as $m \to \infty$. We immediately see that
\begin{align}
\lim_{m \to \infty} Q\left( \frac{\log_2(1+\ssnr|h|^2) - \br_f} {\sqrt{\frac{1}{m} \left( 1- \frac{1}{(\ssnr|h|^2 + 1)^2} \right)} \log_2 e}\right) = \left\{
\begin{array}{ll}
0 & \text{if } \br_f < \log_2(1+\ssnr|h|^2)
\\
1 & \text{if } \br_f > \log_2(1+\ssnr|h|^2)
\end{array}\right.
\end{align}
leading to\footnote{The interchange of the limit and the integral (or equivalently the expectation) can be easily justified by noting the boundedness of the $Q$-function, i.e., $|Q(\cdot)| \le 1$, and invoking the Dominated Convergence Theorem. Additionally, we implicitly assume that the random variable $\log_2(1+\ssnr|h|^2)$ does not have a mass at $\br_f$ and hence $\br_f = \log_2(1+\ssnr|h|^2)$ is a zero-probability event and this event does not affect the expectation.}
\begin{align}
\lim_{m \to \infty} \E_{|h|^2} \left\{Q\left( \frac{\log_2(1+\ssnr|h|^2) - \br_f} {\sqrt{\frac{1}{m} \left( 1- \frac{1}{(\ssnr|h|^2 + 1)^2} \right)} \log_2 e}\right)\right\} = P(\br_f > \log_2(1+\ssnr|h|^2)) \triangleq  P_{\text{out}}.
\end{align}
Therefore, in the limit as $m \to \infty$,
\begin{gather}
R_E(0) \to (1 - P_{\text{out}}) \br_f
\end{gather}
which is defined as the capacity with outage \cite[Section 4.2.3]{goldsmithbook}. Therefore, $R_E(0)$ in (\ref{eq:R_E(0)-fixed}) can be seen as the outage capacity in the finite blocklength regime. Furthermore, $R_E(\theta)$ in (\ref{eq:effth-f}) can be regarded as the generalization of such a throughput measure to the scenario with QoS limitations.

\begin{figure}
\begin{center}
\includegraphics[width=\figsize\textwidth]{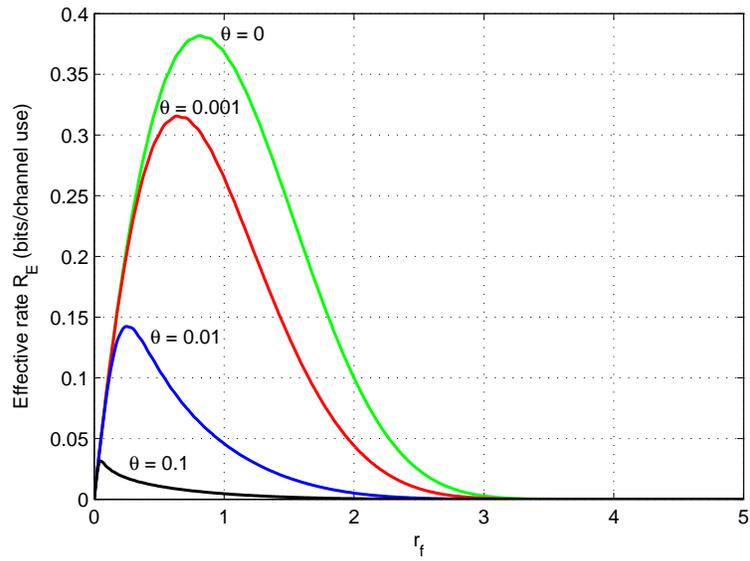}
\caption{Effective rate $R_E$ vs. the fixed transmission rate $\br_f$ in the Rayleigh fading channel. $\tsnr = 0$ dB and the blocklength is $m = 1000$.}\label{fig:R_E-r}
\end{center}
\end{figure}

\begin{figure}
\begin{center}
\includegraphics[width=\figsize\textwidth]{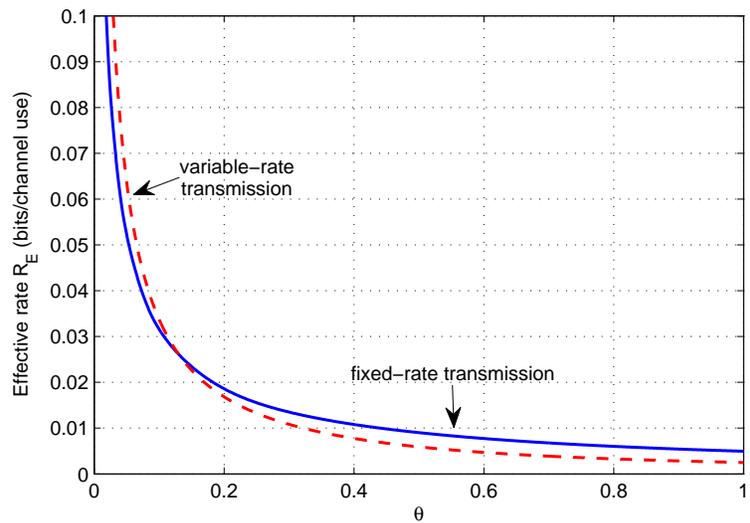}
\caption{The optimal effective rate $R_E$ vs. $\theta$ in the Rayleigh fading channel for both variable-rate and fixed-rate transmissions. $\tsnr = 0$ dB and the blocklength is $m = 1000$.}\label{fig:R_E-theta-fixed}
\end{center}
\end{figure}

\begin{figure}
\begin{center}
\includegraphics[width=\figsize\textwidth]{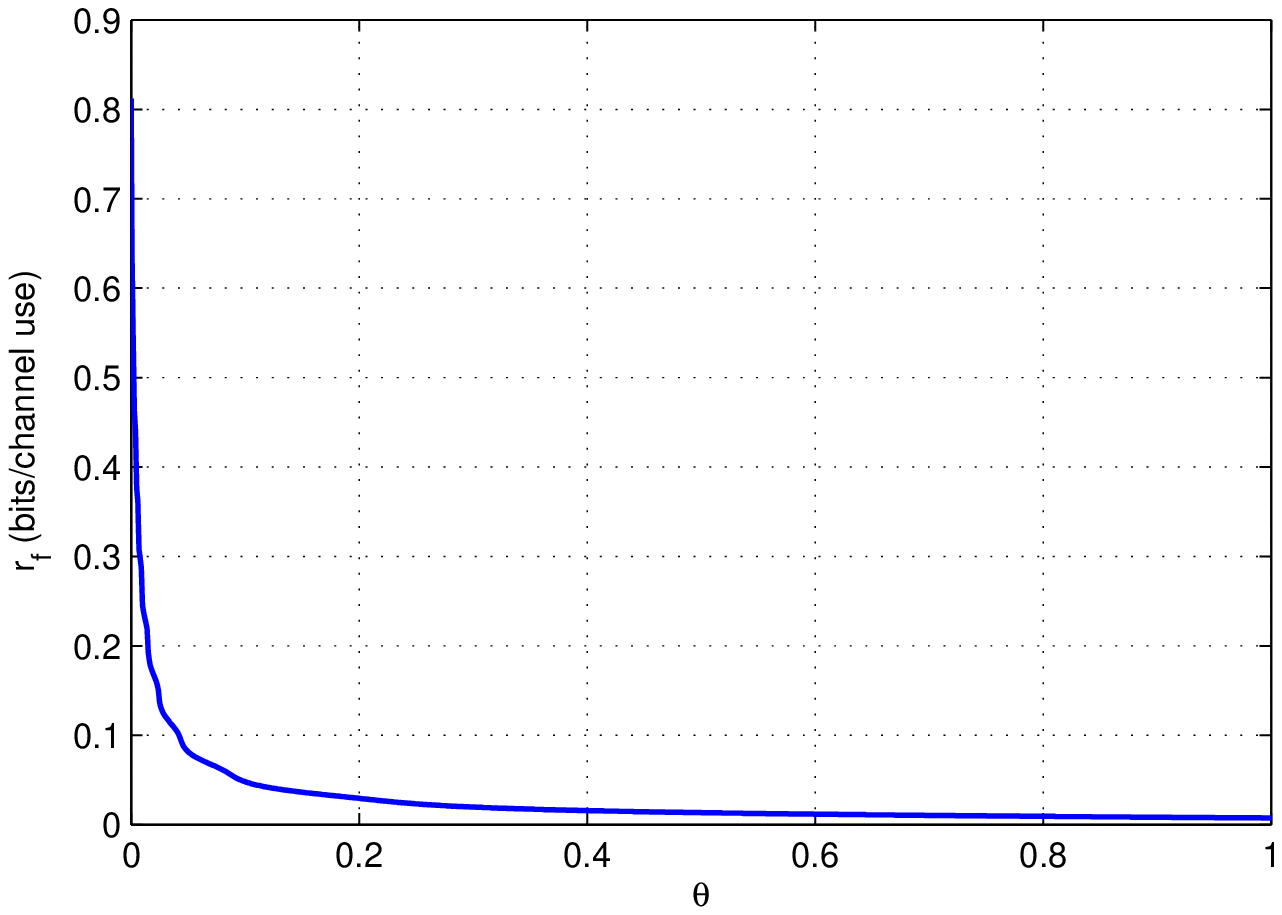}
\caption{The optimal fixed-transmission rate $\br_f$ vs. QoS exponent $\theta$ in the Rayleigh fading channel. $\tsnr = 0$ dB and the blocklength is $m = 1000$.}\label{fig:r_f-theta}
\end{center}
\end{figure}

In Figs. \ref{fig:R_E-r} -- \ref{fig:r_f-theta}, we illustrate the numerical results. In Fig. \ref{fig:R_E-r},  effective rate is given as a  function of the fixed transmission rate $\br_f$. We observe that the effective rate curves are quasiconcave and moreover they are maximized at a unique value  of $\br_f$. We also observe that the maximum value of the effective rate diminishes with increasing $\theta$. This is more clearly seen in Fig. \ref{fig:R_E-theta-fixed} where the optimal effective rates (optimized over $\br_f$) are plotted as a function of $\theta$. In this figure, we have curves for both fixed-rate and variable-rate transmissions. Effective rate for the variable-rate transmission is computed by maximizing (\ref{eq:effthr}) over $\e$. It is interesting to observe that fixed-rate transmissions perform worse than variable-rate transmissions for small values of $\theta$. However, for $\theta > 0.13$, fixed-rate transmissions start outperforming. Hence, for high enough values of $\theta$, fixing the transmission rate and having the error probability vary in each block provide better performance than requiring the error probability to be fixed by varying the rate. Finally, in Fig. \ref{fig:r_f-theta}, we note that as $\theta$ increases, the optimal fixed rate $\br_f$, which maximizes $R_E(\theta)$ in (\ref{eq:effth-f}), diminishes.

\subsection{Sending Independent Messages over Two Parallel Channels} \label{subsec:parallel}

So far, we have assumed that the transmitter sends a single codeword $\x = [\x_r \,\, \x_i]$ of length $2m$ in $m$ channel uses. Another approach is to transmit two independent messages using codewords $\x_r$ and $\x_i$ selected from two independent codebooks. Note that now the codeword length is $m$. These two independent codewords can be seen to be sent through two independent parallel channels:
\begin{align}
\tilde{\y} =   \tilde{\y}_r  + j \tilde{\y}_i = \y e^{-j\theta_h} &= |h| \x + \tilde{\n}
= \underbrace{|h| \x_r + \tilde{\n}_r}_{\text{channel 1}} + j (\underbrace{|h| \x_i + \tilde{\n}_i}_{\text{channel 2}}).
\end{align}
Since the blocklength is $m$ for each codeword, the transmitter sends the information through each channel in the $i^{\text{th}}$ block duration at the following rate with block error probability $\e$:
\begin{align}
\br_{i,p} = &\frac{1}{2}\log_2(1+\ssnr|h_i|^2) - \sqrt{\frac{1}{2m} \left( 1- \frac{1}{(\ssnr|h_i|^2 + 1)^2} \right)} Q^{-1}(\e) \log_2 e \label{eq:codingratef-parallel}
\end{align}
where the subscript $p$ is introduced to differentiate this rate from that in (\ref{eq:codingratef}).
Since errors occur independently in each channel, the service rate (in bits per $m$ channel uses) in each block duration of $m$ channel uses is
\begin{align}
R_i = \left\{
\begin{array}{ll}
0 & \text{with prob. } \e^2
\\
m\br_{i,p} & \text{with prob. } 2\e (1-\e)
\\
2m\br_{i,p} & \text{with prob. } (1-\e)^2
\end{array}\right.. \label{eq:servicerate-parallel}
\end{align}
Effective rate for this service rate  can easily be found as in the proof of Proposition \ref{prop:effrate}, and the proof of the following result is omitted for brevity.
\begin{prop}
When the transmitter sends two independent messages over the independent real and imaginary channels,  the effective rate in bits per channel use at a given $\tsnr$, error probability $\e$, blocklength $m$, and QoS exponent $\theta$ is
\begin{align}
R_E(\theta) &= -\frac{1}{m \theta} \log_e \E_{|h|^2}\left\{\e^2 + 2\e(1-\e)e^{-\theta m\br_p} + (1-\e)^2 e^{-2\theta m\br_p})\right\}
\\
&= -\frac{1}{m \theta} \log_e \E_{|h|^2}\left\{\left(\e + (1-\e)e^{-\theta m\br_p} \right)^2\right\} \label{eq:effthr-parallel}
\end{align}
where $r_p$ is given in (\ref{eq:codingratef-parallel}).
\end{prop}

In this case, it can again be easily shown that the error probability $\e$ that maximizes the effective rate in (\ref{eq:effthr-parallel}) is unique. The following is a corollary to Proposition \ref{prop:convexity}.
\begin{cor} \label{cor:convexity3}
Assume that the values of $m$, $\theta > 0$, and $\tsnr > 0$ are fixed. Then, the function
\begin{gather} \label{eq:functioninsidelog-parallel}
\Psi_p(\e) = \E_{|h|^2}\left\{(\e + (1-\e)e^{-\theta m\br_p})^2\right\}
\end{gather}
is strictly convex in $\e$ and therefore the optimal value of $\e$ that minimizes this function or equivalently maximizes the effective rate in (\ref{eq:effthr-parallel}) is unique.
\end{cor}

\emph{Proof:} See Appendix \ref{app:convexity3}.

\begin{figure}
\begin{center}
\includegraphics[width=\figsize\textwidth]{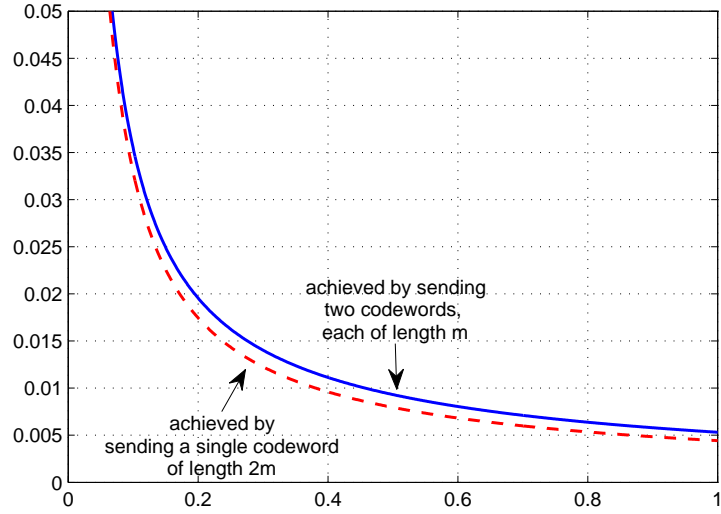}
\caption{The optimal effective rate $R_E$ vs. $\theta$ in the Rayleigh fading channel. $\tsnr = 0$ dB and the blocklength is $m = 1000$. The dashed curve is the effective rate in (\ref{eq:effthr}) maximized over $\e$ and the solid curve is the effective rate in (\ref{eq:effthr-parallel}) maximized over $\e$. }\label{fig:R_E-theta-parallel}
\end{center}
\end{figure}

In the absence of QoS constraints, the effective rate becomes
\begin{align}
R_E(0) &= (1-\e)\E_{|h|^2} \left\{ 2 \br_p \right\}
\\
&=(1-\e)\E_{|h|^2} \left\{\log_2(1+\ssnr|h_i|^2) - \sqrt{\frac{2}{m} \left( 1- \frac{1}{(\ssnr|h_i|^2 + 1)^2} \right)} Q^{-1}(\e) \log_2 e \right\},
\end{align}
which can immediately be seen to be smaller than the effective rate in (\ref{eq:R_E(0)}). Hence, when $\theta = 0$, using two codewords, each of length $m$, provides lower throughput than using a single codeword of length $2m$. Surprisingly, as we observe in Fig. \ref{fig:R_E-theta-parallel}, the throughput achieved by sending two codewords is higher if $\theta$ increases beyond a threshold. Therefore, under strict QoS constraints, sending in each coherence block multiple codewords with shorter lengths may be preferable.

\section{Conclusion} \label{sec:conclusion}

We have analyzed the performance of buffer-constrained wireless systems in the practical scenario in which transmissions are performed using finite blocklength codes with possible decoding errors at the receiver. Employing a recent result on coding rate in the finite blocklength regime, we have determined the effective rate expression as a function of the QoS exponent, coding blocklength, decoding error probability, and signal-to-noise ratio, and characterized the throughput under statistical QoS constraints. We have discussed different transmission strategies. In the case in which the transmission rate is  varied and the error probability is kept fixed across different fading realizations, we have shown that the effective rate is maximized at a unique error probability. This optimal decoding error probability gives us insight on the required reliability of the channel codes. Through numerical results, we have investigated how the optimal effective rate  and optimal error probability vary with the QoS exponent $\theta$. We have also had interesting observations on the performance as a function of the blocklength. We have analyzed the throughput improvements through power adaptation. We have studied the practical scenario in which the transmitter sends the information at a fixed-transmission rate. We have seen that while variable-rate schemes provide higher effective rate at low values of $\theta$, fixed-rate transmissions start performing better as $\theta$ increases. Finally, we have noted that sending multiple codewords with shorter blocklengths in each coherence interval can become a favorable strategy under stringent QoS constraints.

\appendix

\subsection{Proof of Proposition \ref{prop:convexity}} \label{app:convexity}

We first prove the following Lemma.

\begin{lem} \label{lem:f}
For fixed $m$, $\tsnr > 0$, and $|h|^2 > 0$,
\begin{gather}
f(\e) = (1-\e) e^{-\theta m \br}
\end{gather}
 is a strictly convex function of $\e$.
\end{lem}

\emph{Proof}: We first express
\begin{gather}
-\theta m \br = a Q^{-1}(\e) + b
\end{gather}
where, from (\ref{eq:codingratef}),
\begin{gather}
a = \theta \sqrt{m \left( 1- \frac{1}{(\ssnr|h|^2 + 1)^2} \right)} \log e \quad \text{and} \quad
b =  -\theta m\log_2(1+\ssnr|h|^2).
\end{gather}
Note that since $\tsnr >0$, $|h|^2 > 0$ and $\theta >0$, we have $a > 0$. With the above definitions, we can write
\begin{gather}
f(\e) = (1-\e)e^{a \Q + b}.
\end{gather}
The first and second derivatives of $f(\e)$ with respect to $\e$ can easily be found as follows:
\begin{align}
\dot{f}(\e) &= \left[ a (1-\e) \dQ - 1\right] e^{a \Q + b}
\\
\ddot{f}(\e) &= \left[ a (1-\e) (\dQ)^2 - 2 \dQ + (1-\e)\ddQ\right]  a e^{a \Q + b} \label{eq:f''}
\end{align}
where $\dQ$ and $\ddQ$ denote the first and second derivatives, respectively, of $\Q$ with respect to $\e$. Next, we employ several techniques used in \cite[Appendix A]{pengwu} to prove the Lemma. Note that for an invertible and differentiable function $g$, we have $g(g^{-1}(x)) = x$. Taking derivative of both sides of this equality leads us to
\begin{gather}
\dot{g}^{-1}(x) = \frac{1}{\dot{g}(g^{-1}(x))}
\end{gather}
where $\dot{g}^{-1}(x)$ denotes the derivative of $g^{-1}$ with respect to $x$, and $\dot{g}(g^{-1}(x))$ is the derivative of $g$ evaluated at $g^{-1}(x)$. Following this approach and noting that
\begin{gather}
Q(x) = \int_{x}^\infty \frac{1}{\sqrt{2\pi}} \, e^{-t^2/2}\, dt, \quad \text{ and } \quad
\dot{Q}(x) = -\frac{1}{\sqrt{2\pi}} \, e^{-x^2/2},
\end{gather}
we can easily find the following expression:
\begin{gather}
\dQ = -\sqrt{2\pi} e^{\frac{(\Q)^2}{2}}. \label{eq:Q'}
\end{gather}
Note that $\dQ < 0$ for any $0 \le \e \le 1$. Differentiating $\dQ$ with respect to $\e$, we obtain the second derivative as follows:
\begin{gather}
\ddQ = 2\pi \Q \, e^{(\Q)^2} \label{eq:Q''}.
\end{gather}
Next, we consider two cases:

\subsubsection{$\e < 1/2$}

First, we assume that $\e < 1/2$. Under this assumption, we have $\Q > 0$ and hence $\ddQ > 0$. Together with the fact that $\dQ < 0$, we immediately see that
\begin{gather}
\ddot{f}(\e) > 0 \text{ for } \e < 1/2.
\end{gather}

\subsubsection{$\e > 1/2$}

Next, we analyze the case in which $\e > 1/2$ and therefore $\Q < 0$. We concentrate on the term inside the square parentheses in (\ref{eq:f''}). Using (\ref{eq:Q'}) and (\ref{eq:Q''}), and defining $x = \Q$ or equivalently $Q(x) = \e$, we can write
\begin{align}
&a (1-\e) (\dQ)^2 - 2 \dQ + (1-\e)\ddQ \label{eq:e>1/2_0}
\\
&= a (1-\e) 2\pi e^{(\Q)^2} + 2 \sqrt{2\pi} e^{(\Q)^2/2} + (1-\e)2\pi \Q e^{(\Q)^2}
\\
& = a (1-Q(x)) 2\pi e^{x^2} + 2 \sqrt{2\pi} e^{x^2/2} + (1-Q(x))2\pi x e^{x^2}
\\
&= e^{x^2/2} \left( 2\pi (1-Q(x)) (x + a) e^{x^2/2} + 2 \sqrt{2\pi}\right)
\\
&\ge e^{x^2/2} \left( 2\pi (1-Q(x)) x e^{x^2/2} + 2 \sqrt{2\pi}\right) \label{eq:e>1/2_1}
\\
&\ge e^{x^2/2} \left( 2\pi \frac{1}{\sqrt{2\pi}(-x)}e^{-x^2/2} x e^{x^2/2} + 2 \sqrt{2\pi}\right) \label{eq:e>1/2_2}
\\
&\ge e^{x^2/2} \left( -\sqrt{2\pi}+ 2 \sqrt{2\pi}\right) \label{eq:e>1/2_3}
\\
&\ge e^{x^2/2} \left(\sqrt{2\pi}\right) > 0. \label{eq:e>1/2_4}
\end{align}
Above, (\ref{eq:e>1/2_1}) follows from the fact that $a > 0$ and hence $x + a > x$. (\ref{eq:e>1/2_2}) is obtained by using the upper bound,
\begin{gather}
1-Q(x) = Q(-x) < \frac{1}{\sqrt{2\pi}(-x)} e^{-x^2/2} \text{ for } x < 0,
\end{gather}
and recognizing that by our assumption $x = \Q < 0$, and $(1-Q(x))$ is multiplied above by $x < 0$, enabling us to find a lower bound. From the above discussion, we conclude that
\begin{gather}
\ddot{f}(\e) > 0 \text{ for } \e > 1/2.
\end{gather}
Finally, note that when $\e = 1/2$ and hence $\Q = Q^{-1}(1/2) = 0$, we have
\begin{align}
&a (1-\e) (\dQ)^2 - 2 \dQ + (1-\e)\ddQ
\\
&= a (1-\e) 2\pi e^{(\Q)^2} + 2 \sqrt{2\pi} e^{(\Q)^2/2} + (1-\e)2\pi \Q e^{(\Q)^2}
\\
&=a \pi + 2 \sqrt{2\pi} > 0,
\end{align}
and therefore $\ddot{f}(1/2) > 0$. Since $\ddot{f}(\e) > 0$ for all $\e \in [0,1]$, $f(\e)$ is a strictly convex function of $\e$. \hfill $\square$

We now define
\begin{gather}
\psi(\e) = \e + f(\e)  = \e + (1-\e)e^{-\theta m\br}
\end{gather}
which is also strictly convex as it can be immediately seen that $\ddot{\psi}(\e) = \ddot{f}(e) > 0$ for $\tsnr >0$ and $|h|^2 > 0$.
Note that if either $\tsnr = 0$ or $|h|^2 = 0$, the coding rate becomes $\br = 0$, leading to $\psi(\e) = 1 $. Since the nonnegative weighted sum of strictly convex functions is strictly convex \cite{convex} and since the addition of a constant (in the case of $|h|^2 = 0$) does not have an impact on the strict convexity, we immediately conclude that
\begin{gather}
\Psi(\e) = \E_{|h|^2}\left\{ \psi(\e)\right\} = \E_{|h|^2}\left\{\e + (1-\e)e^{-\theta m \br} \right\}
\end{gather}
is strictly convex in $\e$, proving Proposition \ref{prop:convexity}. \hfill $\blacksquare$

\subsection{Proof of Proposition \ref{prop:convexity2}} \label{app:convexity2}

The proof is similar to that of Proposition \ref{prop:convexity} in  Appendix \ref{app:convexity} and will be kept brief. Let's first consider the function
\begin{align}
\phi(\e) &= (1-\e) \br = (1-\e) \left( c_1 - c_2 Q^{-1}(\e)\right)
\end{align}
where we define $c_1 = \log_2(1+\ssnr|h|^2)$ and $c_2 = \sqrt{\frac{1}{m} \left( 1- \frac{1}{(\ssnr|h|^2 + 1)^2} \right)} \log e$. Note that if either $\tsnr = 0$ or $|h|^2 = 0$, then $c_1 = c_2 = 0$ and $\phi(\e) = 0$ for all $\e$. Next, we consider the case in which $\tsnr > 0$ and $|h|^2 > 0$, and therefore $c_1 > 0$ and $c_2 > 0$ \footnote{The strict concavity of the function in the form $(1-\e)(1-\kappa\Q)$ for $\kappa > 0$ is already shown in \cite{pengwu}. We provide a similar proof here for the sake of being complete and keep the discussion brief.}. The second derivative of $\phi(\e)$ with respect to $\e$ is
\begin{gather}
\ddot{\phi}(\e) = 2c_2 \dQ + c_2(\e - 1)\ddQ.
\end{gather}
Using similar arguments as in Appendix \ref{app:convexity}, we  can easily see that for $\e < 1/2$, $\ddot{\phi}(\e) < 0$. For $\e > 1/2$, we can show, employing steps similar to those in (\ref{eq:e>1/2_0})--(\ref{eq:e>1/2_4}), that
\begin{gather}
\ddot{\phi}(\e) < -c_2\sqrt{2\pi} e^{x^2/2} < 0
\end{gather}
where $x = \Q$. When $\e = 1/2$,  we have $\ddot{\phi}(\e) = -2 \sqrt{2\pi} c_2 < 0$. Since $\ddot{\phi}(\e) < 0$ for all $\e$, $\phi(\e)$ is a strictly concave function of $\e$ when $|h|^2 > 0$ and $\tsnr > 0$. As argued similarly in Appendix \ref{app:convexity}, since the nonnegative weighted sum of strictly concave functions is strictly concave \cite{convex} and since the addition of a constant (in the case of $|h|^2 = 0$) does not have an impact on the strict concavity, we conclude that
\begin{gather}
R_E(0) = (1-\e) \E_{|h|^2}\{\br\} = (1-\e) \E_{|h|^2}\{\left( c_1 - c_2 Q^{-1}(\e)\right)\} = \E_{|h|^2}\{ \phi(\e) \}
\end{gather}
is a strictly concave function of $\e$. \hfill $\blacksquare$

\subsection{Proof of Corollary \ref{cor:convexity3}} \label{app:convexity3}

From the proof of Proposition \ref{prop:convexity} in Appendix \ref{app:convexity}, it immediately follows that $\e + (1-\e)e^{-\theta m\br_p}$ is a strictly convex function of $\e$. Then, $(\e + (1-\e)e^{-\theta m\br_p})^2$ is strictly convex due to the facts that $f(x) = x^2$ is a strictly convex and increasing function of $x$ and the composition $f(g(x))$ is strictly convex function when $g(x)$ is a strictly convex function \cite[Section 3.2.4]{convex}. Then, strict convexity of $\E_{|h|^2}\left\{(\e + (1-\e)e^{-\theta m\br_p})^2\right\}$ follows from the arguments employed at the end of Appendix \ref{app:convexity}. \hfill $\blacksquare$

\end{spacing}

\newpage

\end{document}